\documentstyle[12pt,psfig]{l-aa}

\def \rsun {\ifmmode$R$_{\odot}\else R$_{\odot}$\fi}

\def \hcm {\hbox {\ifmmode $ H atoms cm$^{-2}\else H atoms cm$^{-2}$\fi}}
\def \src {3C$\,$273}
\def\approxgt{\mathrel{\hbox{\rlap{\lower.55ex \hbox {$\sim$}}
        \kern-.3em \raise.4ex \hbox{$>$}}}}
\def\approxlt{\mathrel{\hbox{\rlap{\lower.55ex \hbox {$\sim$}}
        \kern-.3em \raise.4ex \hbox{$<$}}}}

\newcommand {\exosat} {{EXOSAT}}
\newcommand {\asca} {ASCA}

\newcommand {\sax} {BeppoSAX}

\newcommand {\etal} {et~al.}

\newcommand {\ergs} {erg~s$^{-1}$}

\psfigurepath{/usr4/users/aorr/mcg-6-30-15}
\begin{document}

\thesaurus{ (03.09.3; 11.17.1; 11.17.4; 13.25.3)}

\title{Quasi-simultaneous observations of \src\ by \sax\ and \asca}

\author{
Astrid Orr\inst{1}
\and T. Yaqoob\inst{2}\thanks{\hspace{-0.3cm}Universities Space Research Association}
\and  A.N.~Parmar\inst{1} 
\and L. Piro\inst{3}
\and N.E. White\inst{2} 
\and P. Grandi\inst{4}}
\institute{Astrophysics Division, Space Science Department of ESA, 
ESTEC, P.O. Box 299, 2200 AG Noordwijk, The Netherlands
\and
Laboratory for High Energy Astrophysics, NASA Goddard
Space Flight Center, Greenbelt, MD 20771, USA
\and
Istituto di Astrofisica Spaziale, CNR, Via Fosso del Cavaliere, 
I-00133 Roma,  Italy
\and
Istituto di Astrofisica Spaziale, C.N.R., Via Enrico Fermi, I-00044 Frascati,
Italy
%\and
%\sax\ Data Center, c/o Nuova Telespazio, via Corcolle 19, I-00131 Roma, 
%Italy
%\and Unita' G.I.F.C.O.-CNR, via Archirafi 36, I-90123 Palermo, Italy
}
\date{Received ; accepted}
\offprints{A. Orr: aorr@astro.estec.esa.nl}
\maketitle
\markboth{A. Orr et al.: X-ray observations of \src}{A. Orr et al.:
X-ray observations of \src}

\begin{abstract}
We report the results of quasi-simultaneous observations of the 
bright quasar \src\ using the \sax\ Low-Energy Concentrator Spectrometer
(LECS) and the \asca\ Solid-State Imaging Spectrometer (SIS).
These observations are part of an intercalibration programme and 
demonstrate  good agreement between
both instruments in the common energy range 0.5-10~keV. In particular,
the absorption feature discovered by BeppoSAX/LECS at 
$\sim$0.5~keV is consistently seen by ASCA/SIS.
We present the most accurate measurement yet of the spectral shape of this 
absorption feature and provide constraints on its physical
interpretation. A self-consistent warm absorber model
in photoionization equilibrium only gives an adequate spectral fit 
if  a very high velocity inflow is allowed for the warm absorber  
(v$\sim$0.36 c). The ionization parameter and the column density
of the warm absorber  are then $\sim$1 and  
$\sim$1.3 $\times 10^{21}$ cm$^{-2}$, respectively.     
If the differences in best-fit neutral absorption column density 
between the LECS and 
each SIS are taken as estimates of the systematic spectral uncertainties, 
then these are $<$0.8$\times 10^{20}$ and $<$2.0$\times 10^{20}$ cm$^{-2}$ 
(at 90\% confidence) for SIS0 and SIS1, respectively.

\end{abstract}

\keywords{Instrumentation: miscellaneous -- Quasars: individual: \src~-- 
Quasars: absorption lines -- X-rays: general}

\section{Introduction}
\label{sec:introduction}

\src\ is a high luminosity radio-loud quasar at a redshift of z=0.158.
Previous X-ray observations of \src\ using HEAO 1 A2,
\exosat\ and {\it Ginga} showed a featureless power-law continuum with 
a photon index, $\Gamma$, that
varies between 1.3 and 1.6 (Worrall \etal\ 1979; Turner \etal\ 1990, 1991;
Williams \etal\ 1992). The 2--10~keV luminosity varies in the range
0.7--2.0$\times 10^{46}$ \ergs (Turner \etal\ 1990). 
EXOSAT observations revealed a soft excess below 1~keV 
(Courvoisier \etal\ 1987),
which has been confirmed by subsequent ROSAT observations 
(Staubert 1992; Staubert \etal\ 1992). 
This soft excess is also visible in an \asca\ Performance Verification
observation of \src, but at a reduced level compared to that observed
by ROSAT (Yaqoob et al. 1994).
Both {\it Ginga} (Williams \etal\ 1992) and ASCA observations 
(Cappi \& Matsuoka 1997)
indicate the presence of a weak narrow Fe K$\alpha$ emission line 
(equivalent width, EW, $\sim50$~eV) during low X-ray flux states
and there is recent evidence for a strong broad Fe K$\alpha$ emission line
(EW $\approxgt$100~eV, Yaqoob et al. in preparation).
The \sax\ Science Verification Phase (SVP) observation also took place during 
a low state. The SVP spectrum can
be  represented by a single power-law model between 1 and 200~keV with  
a weak narrow Fe emission feature at a rest frame energy of 6.4~keV 
with an equivalent width EW of 30~eV (Grandi et al. 1997). Below 1~keV
the BeppoSAX LECS detected a soft excess and led to the discovery of an 
absorption feature in this source at $\sim$0.5~keV (Grandi et al. 1997,
Fossati \& Haardt 1997). 

\src\ was observed quasi-simultaneously by \sax, \asca, 
and the Rossi X-ray Timing Explorer (XTE) as part of a dedicated program to 
cross-calibrate the instruments on the three satellites.
\src\ was chosen since it is visible at the lowest and highest energies
observable by the above satellites and because of the supposed 
simplicity of the underlying spectrum.
Here we describe one aspect of the program, the 
intercalibration of the \sax\ LECS and \asca\  SIS 
instruments. Both instruments detect similar low-energy 
spectral structure from \src, indicating a  good relative calibration.

\section{The instruments}
\label{instruments}

The payload of the Italian-Dutch satellite \sax\ includes four identical 
X-ray mirrors with imaging gas scintillation proportional counters 
(GSPCs) as detectors (Boella et al. 1997a). Amongst these the LECS is 
sensitive from 0.1--10\,keV (Parmar et al. 1997). 
The other detectors operate in the energy
range 1.3--10\,keV and are referred to as the Medium-Energy 
Concentrator Spectrometer (MECS; Boella et al. 1997b). 

The LECS achieves an extended low-energy response by utilizing a
driftless gas cell and an ultra-thin entrance window. 
The energy resolution of the LECS is 32\% full width at half-maximum 
at 0.28~keV, 16\% at 1.5~keV and  8.8\% at 6~keV. 
For an on-axis source, the effective area of the LECS peaks
at $\sim$50~cm$^2$ at an energy of $\sim$1.5~keV and  
the area is $\sim$11~cm$^2$ directly below the carbon edge at 0.28~keV.
A study of a large sample of LECS AGN spectra below 2~keV (Orr et al. 1998) 
has shown that the maximum amplitude of fit residuals due to instrumental 
effects is  $\pm$10\% of the folded model count rate
in the energy ranges 0.1--0.4~keV and 0.5--2~keV. Between 0.4--0.5~keV,
where the effective area is very low,
effects up to a 15\% level are present with the current LECS response
({\tt LEMAT 3.5.3}).

\asca\ has four identical X-ray telescopes. The corresponding focal plane 
detectors are two Solid-state Imaging Spectrometers (SIS) and two Gas Imaging
Spectrometers (GIS) (Tanaka et al. 1994; Serlemitsos et al. 1995).
The SIS is sensitive in the energy range 
0.4--10\,keV. Originally, the SIS energy resolution was 
5\% at 1.5~keV and 2\% at 6\,keV, but due to radiation damage 
it was reduced to  3.6\% and 3.9\% at 6~keV for SIS0 and SIS1, respectively,
at the time of the observation (see Dotani et al. 1997).

The low-energy responses of the SIS and the LECS are difficult to 
calibrate in-flight. The Crab Nebula, which is a favored calibration source 
in X-ray astronomy, is not detected by the LECS  
$\approxlt$0.5~keV due to interstellar absorption, and observation of such 
a bright source  
is not possible with the SIS because of multiple events occurring in 
single pixels.
For this reason \src\
was used to cross-calibrate the SIS and GIS above 3~keV (Dotani et al. 1996).
The GIS can observe the Crab, so this source was
used to calibrate the overall response of the GIS detector and to verify the
effective area of the X-ray telescope.

\section{Observations and data reduction}
\label{observations}

\src\ was observed by \sax~between 1996 July 18--21 during the SVP 
and by \asca\ between 1996 July 16--18 (Table 1). 
Good LECS data were selected from intervals when the 
minimum elevation angle above the Earth's limb was $>$4$^{\circ}$ and when the 
instrument parameters were nominal.
Due to an instrument anomaly, LECS data between 
1996 July 19 11:23 and July 20 13:06~UTC were lost.
Since the LECS was only operated 
during satellite night-time, this gave a total on-source exposure of
11.5~ks. 
LECS data were processed using {\tt SAXLEDAS 1.7.0} 
(Lammers 1997).  

A spectrum was extracted at the source centroid using the standard
extraction radius of 8$'$. 
Background subtraction was
performed using standard blank field exposures, but is not critical
for such a bright source. 
The \src\ count rate above background is 
$0.57$~s$^{-1}$ and the background in the 0.1--10~keV range
is $<$4\% of the total count rate. 
\src\ was located $\sim$2$'$ off-axis in the LECS FOV. This position is close
to a strongback rib, which reduces the overall source count rate.
Since the LECS response is dependent on both source position within the FOV 
and extraction radius, the appropriate matrix was created using 
{\tt LEMAT 3.5.3}.

%------------------------------table-1--------------------------------------
 \begin{table}
\caption[]{\src\ observation log}
\begin{flushleft}
\begin{tabular}{lllll}
\hline\noalign{\smallskip}
Satellite &Inst. & Start time & End time & Exp. \\
          &           &(1996)   (UTC)    & (UTC)    & (ks)     \\
\noalign {\smallskip}
\hline\noalign {\smallskip}
\sax  & LECS & July 18~00:41 & July 21~08:35 & 11.6 \\
\asca & SIS0 & July 16~19:52 & July 17~14:54 & 42.0 \\
      & SIS1 & July 16~19.52 & July 18~00:57 & 41.9 \\
\noalign {\smallskip}
\hline
\end{tabular}
\end{flushleft}
\label{obs_log}
\end{table}
%------------------------------table-1--------------------------------------

The two
SIS instruments observed \src\ on the default chips, in 1-CCD FAINT mode.
The data were cleaned by excluding
South Atlantic Anomaly passages (SAA), applying the standard algorithm
for removing hot and flickering pixels, requiring an Earth elevation angle
greater than $5^{\circ}$ and a magnetic cut-off rigidity greater than 
7~GeV~c$^{-1}$.
Data were also rejected within 50~s of the satellite exiting the SAA
and within 50~s of day/night transitions. Standard upper limits of 100 were
also applied to the `S0\_PIXL1' and `S1\_PIXL3' parameters in order
to eliminate contamination due to light leaks in the SIS. 

On-source data were extracted using circles of radius 4$'$ centered on the
source. Background spectra with sufficient signal-to-noise
were extracted from off-source regions of the chips. 
The count rates in SIS0 and 
SIS1 are 2.36 and 1.88 s$^{-1}$, respectively, with the background
constituting $<$4\% of the total counts in the 0.5--10~keV band.
The response matrices for the SIS were generated using
{\tt sisrmg 1.1}, with the CTI tables issued on 1997 March 11.
The XRT responses were generated with {\tt ascarf 2.64}, with
both of the empirical energy-dependent effective area modification
factors implemented.
In the following, all spectra are re-binned to have $>$20
counts in each bin allowing the use of the $\chi^2$ statistic.
Uncertainties are quoted as 90\% confidence 
intervals for one parameter of interest. 

\section {Spectral fits}
\label{spectrum}
Fits were performed on the SIS and LECS data in different energy bands
in order to examine whether a single absorbed power-law provides an adequate 
description of the entire spectrum. 
A model without an Fe K$\alpha$ emission line was used in order to 
have a simple parameterization. 
Furthermore, the presence of the line does not affect the conclusions
of the present analysis. 
The photoelectric absorption coefficients of 
Morrison \& McCammon (1983) and the abundances of Anders \& Grevesse 
(1989) were used throughout the fitting.

\subsection{SIS-only fits}
\label{sisonly}

In the 2--10\,keV energy range independent fits to the SIS0 and 
SIS1 spectra using a simple power-law 
with neutral absorption fixed at the galactic value, ${\rm N_{H}(Gal)} = 
1.84 \times 10^{20}$ cm$^{-2}$ (Stark et al. 1992),
give photon indices ${\rm \Gamma_{SIS0}}$ and ${\rm \Gamma_{SIS1}}$
of 1.59$\pm 0.03$. 
The fit statistics are better for SIS1 than
for SIS0 ($\chi^2_{\rm SIS0}$ = 251.5 for 208 degrees of freedom, dof; 
$\chi^2_{\rm SIS1}$ = 186.6 for 199 dof). 
The indices agree well with the values
obtained for the same model using the near-simultaneous 
XTE/PCA and BeppoSAX/MECS observations. $\Gamma_{\rm XTE}$(2--10\,keV) = 
1.61$\pm 0.01$ (Yaqoob et al. in preparation) and
${\rm \Gamma_{MECS}}$(1.5--10~keV) = 1.57$\pm 0.01$ (Grandi et al. 1997), using
the slightly lower value of ${\rm N_{H}(Gal)} = 
1.68 \times 10^{20}$ cm$^{-2}$ from Savage et al. (1993).

When the low energy data down to 0.5\,keV are included and 
excess neutral absorption is allowed, the simple power-law fits 
result in basically the same slopes and quality of fit 
(${\rm \Gamma_{SIS0,1}}$(0.5--10\,keV) = 1.63$\pm 0.02$). 
However, excess ${\rm N_{H}}$ of (3.3$\pm0.9$)$\times 10^{20}$ cm$^{-2}$ 
and (5.5$\pm1.0$) $\times 
10^{20}$ cm$^{-2}$ are required for SIS0 and SIS1, respectively.

\subsection{LECS-only fits}
\label{lecsonly}
A 2--10\,keV power-law fit
with galactic absorption gives a marginally flatter slope in the LECS 
than measured by the SIS, ${\rm \Gamma_{LECS}}$(2--10~keV) =  
$1.51 \pm 0.07$ ($\chi^2$ = 140.2 for 135 dof).
Including the data at lower energies (0.5--10\,keV) and allowing for excess
neutral absorption, ${\rm N_{xs}}$, also gives a good fit 
($\chi^2$ of 232.2 for 242 dof). The spectral slope is similar to above,
${\rm \Gamma_{LECS}}$(0.5--10~keV)$ = 1.57 \pm 0.05$,
and the excess column density increases to a value 
compatible with  that found by  the SIS in this energy range. However,
the uncertainty on  excess absorption is large,
${\rm N_{H}} = (2.7 \pm^{2.7}_{2.6}) \times 10^{20}$ cm$^{-2}$. 
When data down to 0.1 keV is included, and using only galactic neutral
absorption, one obtains an unacceptable fit with significant residuals 
below 0.8\,keV and a $\chi^2$ of 313.5 for 263 dof. 
If the total absorption is left as a free
parameter it decreases to $(0.7\pm {0.2})\times 10^{20}$ cm$^{-2}$
and the fit statistics improve by $\Delta(\chi^2)$ =  52.3, as the fit 
tries to account for the increase in flux $\approxlt$0.4\,keV. 

In summary, across their common energy range (0.5--10\,keV)
both the LECS and the SIS data show that the spectrum of 
\src\ deviates from a single power-law at lower energies and 
requires excess neutral absorption. 
In this energy range the best-fit values of $\Gamma$ and ${\rm N_{H}}$ are 
closely coupled. 
The simple power-law model with galactic and excess
neutral absorption and  provides a
satisfactory, if not physical, description of both the LECS and the 
\asca\ data between 0.5--10\,keV. However, it becomes unacceptable for fits
extending below 0.5~keV. 
Such behavior is suggestive of spectral structure. In fact,
structured residuals at $\sim$0.5--0.6\,keV have been reported by 
several workers in the case of SIS observations of 
\src~ (Yaqoob et al. 1994) and other AGN, 
leading to speculation about systematic uncertainty of the SIS response
(Dotani et al. 1996; Cappi \& Matsuoka 1997).
Moreover, a previous analysis of the LECS SVP data (Grandi \etal\ 1997)
revealed for the first time in this source the presence of an absorption 
feature at $\sim$0.5~keV.

\subsection{Estimates of the relative SIS/LECS systematic deviation}
\label{systematic}
In order to verify whether similar low energy structure is present
in both LECS and SIS data for \src\ simultaneous 
SIS0/LECS and SIS1/LECS spectral fits were performed 
in the energy range 0.6--4\,keV. 
It is assumed that any relative systematic deviation between the SIS and 
the LECS in the range
0.6--4\,keV can be represented by an effective column density. 
Data below 0.6\,keV were not used  because the SIS detection efficiency
is poorly understood here, probably due to the effects of an O edge  
at 0.53\,keV (Dotani et al. 1996). 
By restricting the energy range to below 4~keV relatively more statistical 
weight is given to the low energy bins which are affected by the spectral 
complexity.

The model combines a normalization factor, a power-law and three
neutral absorption components: one fixed at the galactic value of
${\rm N_{H}(Gal)} =  1.84 \times 10^{20}$ cm$^{-2}$, a second column 
density common to both LECS/SIS0 or LECS/SIS1 and a third component which is 
zero for the LECS and allowed to vary for the SIS. This third component 
characterizes the systematic deviation relative to the LECS. 
The choice of the model is based on the results of Sects. \ref{sisonly} and 
\ref{lecsonly}.
%------------------------------figure-1--------------------------------------
\begin{figure}[htb]
\hbox{\psfig{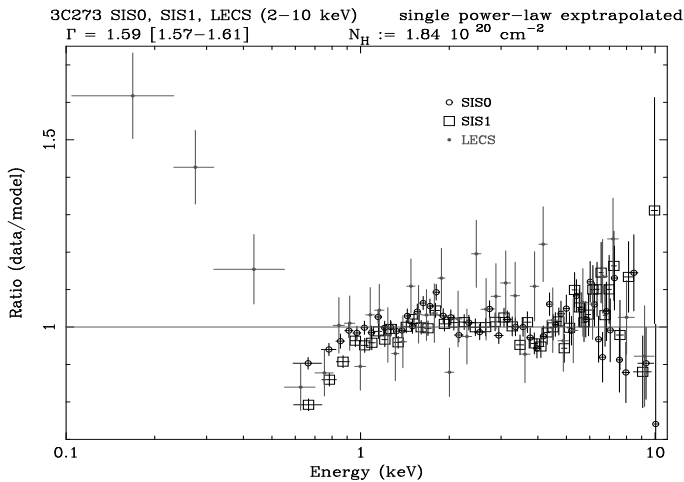} }
\caption{\protect \small Data-to-model ratio of the 
ASCA--BeppoSAX spectrum of \src~:
SIS0, SIS1 and LECS data between 
2--10~keV are fit with a single power-law and neutral absorption
fixed at ${\rm N_{H}}$(Gal). 
The resulting model is extrapolated to lower energies showing evidence for 
complex spectral structure $\approxlt$1~keV.
The data have been re-binned for the plot}
\label{sislecs}
\end{figure}
%------------------------------figure-1--------------------------------------

%------------------------------table-2--------------------------------------
\begin{table}
\caption[]{Simultaneous SIS/LECS spectral fits in the range 0.6--4\,keV. The 
fit model is described in the text. 
${\rm N_{xs}}$ and ${\rm N_{sys}}$ are in units 
of $10^{20}$ cm$^{-2}$. The normalization is SIS/LECS }
%\begin{flushleft}
\scriptsize{
\begin{tabular}{llllll}
\hline\noalign{\smallskip}
SIS & $\Gamma$ & ${\rm N_{xs}}$ & ${\rm N_{sys}}$ & Norm & 
$\chi^2$ (dof) \\
\noalign {\smallskip}
\hline\noalign {\smallskip}
0 & $1.68\pm 0.03$ & 4.7$\pm$1.1 & $0.0 \pm _{0.0}^{0.8}$ 
& $1.09 \pm 0.03$  & 319.1 (282) \\
1 & $1.68\pm 0.03$ &  $6.7 \pm _{2.1}^{1.0}$ & 
$0.0 \pm _{0.0}^{2.0}$ & $1.05 \pm 0.04$ & 284.1 (282) \\
\noalign {\smallskip}
\hline
\end{tabular}
}
%\end{flushleft}
\label{tabsystem}
\end{table}
%------------------------------table-2--------------------------------------
The model provides a good fit to the data.
The result (Table \ref{tabsystem}) is that the relative SIS/LECS 
systematic deviation ${\rm N_{sys}}$  in the common energy range 0.6--4\,keV 
is consistent with zero. 
The average data-to-model ratio for the fits in Table \ref{tabsystem} and 
for either SIS0 or SIS1 has a standard deviation of 6\% ({\it rms}). 
The corresponding standard deviation for the LECS is 23\%. 
When computed for the SIS in the energy range 0.6--1\,keV, the 
standard deviation  is even smaller
(4\%), whereas the LECS standard deviation decreases to 19\%.

These results remain valid even with a more complex fit model. 
For instance, a model was considered with the same components
as the previous one, but extending down to 0.5~keV and
including an absorption edge at E$_{\rm rest}\sim$0.6\,keV. The justification for 
such a model is given in the following section.
Figure \ref{sislecs} illustrates the good agreement between LECS and SIS 
at low energies. Here
SIS0 and SIS1 data from 2--10~keV and BeppoSAX/LECS data  between 
2--4~keV are fit with a single power-law and neutral absorption
fixed at ${\rm N_{H}(Gal)} = 1.84 \times 10^{20}$\,cm$^{-2}$. 
The resulting model
($\Gamma = 1.59 \pm 0.02$ with a $\chi^2$ 589.5.5 for 545 dof) 
is extrapolated to lower energies. 

\subsection{Fitting the absorption feature: 0.1-10\,keV fits}
\label{absfits} 
 
%------------------------------table-3--------------------------------------
\begin{table}
\caption[]{Simultaneous spectral fits LECS: 0.1--10\,keV; SIS0, 
SIS1 0.5--10\,keV. All models include galactic neutral absorption with 
${\rm N_{H}(Gal)} = 1.84 \times 10^{20}$\,cm$^{-2}$ 
and all but model 4 allow for 
free excess absorption. Model 3  is composed of two power-laws and a partial 
covering fraction absorption. Models 4, 5, 10 and 11 include broken power-laws.
Models 12, 13 and 14 include blackbody spectra}
\begin{tabular}{lr}
\hline\noalign{\smallskip}
n Model & $\chi^2$ (dof) \\
\noalign {\smallskip}
\hline\noalign {\smallskip}
1 PL (power-law) & 1144.2 (775) \\
2 PL + PL & 874.1 (773)\\
3 (PL + PL) $\times$ part. CF abs. & 874.1 (771) \\
4 BPL (${\rm N_{H}(Gal)}$) & 1116.7 (774) \\
5 BPL  & 880.3 (773) \\
6 PL $\times$ edge & 881.7 (773) \\
7 (PL + PL) $\times$ edge & 850.0 (771) \\
8 (PL + PL) $\times$ notch & 858.6 (770) \\
9 (PL + PL) $\times$ inv. Gauss. & 859.4 (770)\\
10 BPL $\times$ edge & 850.7 (771)\\
11 BPL $\times$ notch & 865.6 (770)\\
12 (BB + PL) & 878.9 (773)\\
13 (BB + PL) $\times$ edge & 851.9 (771)\\
14 (BB + PL) $\times$ notch & 861.3 (770)\\  
\noalign {\smallskip}
\hline
\end{tabular}
\label{trials}
\end{table}
%------------------------------table-3--------------------------------------
%------------------------------table-4--------------------------------------
\begin{table}
\caption[]{Parameters of best-fit models 7, 10 and 13 of Table~\ref{trials}. 
The edge energy is in the source rest frame  }
\scriptsize{
\begin{tabular}{lllllll}
\hline\noalign{\smallskip}
n &${\rm \Gamma_{soft}}$ & ${\rm \Gamma_{hard}}$ & E${\rm _{edge}}$ 
& $\tau$ & ${\rm N_{xs}}$ \\
 & T$_{\rm soft}$ eV& & (keV) & & ($10^{20}$cm$^{-2}$)\\
\noalign {\smallskip}
\hline\noalign {\smallskip}
7 & $ 3.1 \pm _{0.9}^{3.2}$ &  $1.60 \pm _{0.11}^{0.03}$ & $0.58 \pm 0.01$ & $0.70 \pm _{0.21}^{0.12}$ &
$\leq 1.1$\\
10 & $2.0 \pm _{0.1}^{0.5}$ & $1.62\pm 0.01$ & $0.58\pm 0.01$ & $0.60
\pm _{0.13}^{0.09}$ & $\leq 0.1$\\
13 & $43 \pm _{20}^{48}$ & $1.62\pm 0.01 $ & $0.59\pm _{0.02}^{0.01} $ & $0.54
\pm _{0.12}^{0.32}$ & $\leq 0.9$\\
\noalign {\smallskip}
\hline
\end{tabular}
}
%\end{flushleft}
\label{fitpar}
\end{table}
%------------------------------table-4--------------------------------------
%------------------------------figure-2--------------------------------------
\begin{figure}[htb]
\hbox{\psfig{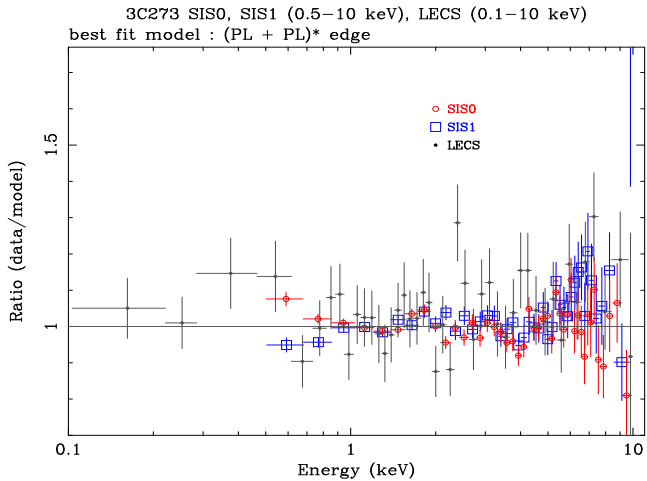} }
\caption{\protect \small Model 7 of Table \ref{trials}. 
``Data-to-model'' ratio of the best fit to SIS (0.5--10~keV) and LECS 
(0.1--10~keV) data: (PL + PL) $\times$
edge. The fit parameters are listed in Sect.~\ref{absfits}. 
The data have been re-binned for the plot}
\label{bestfit}
\end{figure}
%------------------------------figure-2--------------------------------------

The results of Sect.~\ref{lecsonly} reveal that complex spectral
structure is present below $\sim$1\,keV. Further fits have
been performed between 0.1--10~keV (LECS) and 0.5--10~keV (SIS0, SIS1) 
in order to measure the spectral shape of this structure.
The results of various simultaneous LECS/SIS0/SIS1 fits are 
listed in Table 
\ref{trials}. The fit models include normalization factors for each 
instrument. The factors were left as free parameters, 
but remained very stable from one model to the next, the average values 
with respect to the LECS being:
SIS0/LECS = 1.07, SIS1/LECS = 1.04 with a typical uncertainty of $\pm 0.03$. 

A single power-law model is unacceptable. 
The improvement in $\chi^2$ fit statistics brought by adding a ``soft excess''
component is significant at $>$99.9\% confidence using the F-statistic
(see models 2, 5 and 12 versus model 1), with the two power-law model (2) 
giving slightly better results than the broken power-law or the blackbody plus
power-law model. However, the fits remain unacceptable, with large
residuals below 1~keV.
The addition of a narrow Fe K$\alpha$ emission line at $\sim$6.4~keV 
(EW $\sim$24 eV) brings
no significant improvement to the quality of fit ($<$95\% confidence level
with the F-statistic) and is therefore omitted in the present study.
Furthermore, the presence of a broad Fe K$\alpha$ emission component does not
affect the results of the present study.
 
Models including an absorption feature do bring  significant improvement
to the fit statistics. 
The ``best fit'' is obtained by including an absorption feature in the 
form of an edge (models 7, 10 and 13 of Table \ref{trials}, 
see Fig.~\ref{bestfit}). 
It is not possible to discriminate between the soft continuum
components (i.e. two power-laws, a broken power-law or 
by a blackbody and power-law) when an absorption edge is included, since
the fits statistics are very similar.
The best fit parameters are given in Table \ref{fitpar}. 
In broken power-law models the energy of the break, when left as a free
parameter (models 5, 10 and 11) tends to values close to that of the
absorption feature, thereby indicating that such a continuum contributes in 
part to the fit of the absorption feature.

The observed 2--10~keV flux with model 7 and the LECS normalization is 
$6.8 \times 10^{-11}$ ergs s$^{-1}$ cm$^{-2}$.
The 0.15--2~keV unabsorbed fluxes in the soft component are 1.0 and $0.5 
\times 10^{-11}$ ergs s$^{-1}$ cm$^{-2}$ with models 7 and 13 respectively. 
These soft excess values are somewhat lower than those
measured by ROSAT/PSPC (Walter et al. 1994) in the same spectral range
during the ROSAT all sky survey (RASS), 
i.e. 4.7 and $4.8 \times 10^{-11}$ ergs s$^{-1}$ cm$^{-2}$ with  two power-laws
and a blackbody plus power-law fit respectively. 
On the other hand, the flux at 2~keV observed by BeppoSAX/ASCA, 
$\nu {\rm F_{2keV}(BeppoSAX/ASCA)} \sim 4\times 10^{-3}$ 
photons~keV$^{-1}$ s$^{-1}$ cm$^{-2}$, is comparable to the RASS flux,
$\nu {\rm F_{2keV}(RASS)}= (2.86 \pm 0.37)\times 10^{-3}$ 
photons~keV$^{-1}$ s$^{-1}$ cm$^{-2}$. Leach et al. (1995) have shown that 
the 1.5--2.4~keV and 0.1--0.3~keV count rates of \src~are both highly variable
but uncorrelated.
 
Slightly worse fits are obtained when the absorption feature is modeled  
by a notch or by a Gaussian line (models 8, 9, 11 and 14).
The Gaussian absorption line (model 9) has a line energy of E$_{\rm rest}=
0.68 \pm{0.04}$~keV, a line width $\sigma= 114\pm_{43}^{99}$ eV and EW = 
$33.3 \pm_{20.6}^{14.7}$ eV (the uncertainty intervals for E$_{\rm rest}$
and EW are calculated with $\sigma$ fixed at 114 eV). The rest frame energy 
of the line is compatible with resonant absorption from 
O{\sc viii} Ly$\alpha$ at E$_{rest} = 0.654$~keV.
However, significant structure remains in the fit residuals at the 
energy of the absorption feature when it is modeled by  a 
Gaussian line or a notch.

\subsection{Fitting the absorption feature: photoionization models}
\label{photfits} 

To understand the physics of the absorption feature, we have calculated a
two-dimensional grid of photoionization models using the code 
CLOUDY (Ferland 1996). 
The two grid parameters are U, the dimensionless ratio of hydrogen ionizing
photon density to total hydrogen density and N$_{\rm W}$, the total hydrogen
column density of of the ionized material in units of cm$^{-2}$.
N$_{\rm W}$ determines the amplitude of a given absorption feature, 
whereas the parameter U dictates the abundances of different
ions and therefore also the presence of absorption features at
a given energy. 

The following assumptions have been made. The warm absorber is in 
photoionization equilibrium with a central ionizing continuum.
The ionizing continuum (here the ``standard AGN'' continuum in CLOUDY) 
consists of an ``X-ray'' power-law of photon index 
$\Gamma_{\rm X} = 1.6$
from 1.36 eV to 100~keV, above which the continuum falls off as E$^{-3}$. A
``Blue Bump'' power-law is added to the ``X-ray'' power-law, with   
$\Gamma_{\rm BB} = 1.5$ and an exponential cut-off corresponding to a
``Blue Bump'' temperature of $1.5 \times 10^4$ K. A typical X-ray to UV
ratio was chosen so that $\Gamma_{\rm ox} = 2.4$, where $\Gamma_{\rm ox}$
is the spectral index of a single power-law connecting the observed fluxes
at  2~keV and 2500~\AA~(see CLOUDY manual).
Solar composition (Grevesse \& Anders 1989) and a constant density 
are assumed (with a hydrogen density n$_{\rm H} = 10^6$ cm$^{-3}$) within the
entire single-zone warm absorber. The fits were performed between 
0.1--5~keV in order to give relatively more statistical weight to the low 
energy data, 
neutral absorption was fixed at the galactic value and the
instrumental normalization factors were fixed at the values listed in
Sect.~\ref{absfits}.

Both stationary and red-shifted warm absorbers were considered.
As noted by Grandi et al. (1997) the stationary case gives a very poor 
fit to the data. This is because
the bulk of the model absorption is caused at E$>$0.7~keV, by the 
O{\sc vii} and O{\sc viii} edges, which is at significantly 
higher energy than observed.
When the red-shift of the warm absorber is left as a free parameter, the best
fit parameters are z$=0.58 \pm _{0.09}^{0.02}$ with respect to the observer's 
frame (i.e. z$\sim$0.36 with respect to the rest frame of \src),
log U$=0.0\pm _{0.21}^{0.17}$ and log N$_{\rm W} =  21.13 \pm_{0.09}^{0.06}$, 
with $\chi^2 =$ 667 (for 516 dof). 
When the neutral absorption is left as a free parameter the statistics
improve ($\chi^2 = $642.2 for 515 dof). 
These results can be compared to the statistics obtained in the same 
energy range with a single power-law and free neutral absorption 
($\chi^2 =$ 881.3 for 515 dof) or a broken power-law and free
neutral absorption ($\chi^2 = 601.3$ for 513 dof).

\section {Discussion}
\label{subsec:discussion}

Both separate and simultaneous 
SIS and LECS fits make it possible to constrain the spectral shape of \src~and 
to measure the relative systematic deviation between the SIS and the LECS.
A power-law with galactic neutral absorption gives an adequate description of
both the SIS and the LECS data in the range 2--10~keV. The SIS power-law 
slope ($\Gamma = 1.59 \pm 0.03$) agrees well with near simultaneous
values from XTE and MECS (Sect.~\ref{sisonly}). 
%Because of its off-axis exposure and penetration effects 
%(see Sect.~\ref{lecsonly}), 
The 2--10~keV LECS spectrum of \src~appears marginally flatter than the SIS 
spectrum ($\Delta (\Gamma) = 0.08 \pm 0.08$). 
Between 0.5--2~keV, the SIS and the LECS spectra show evidence for a flux 
deficit at the lowest energies which can be fit with excess neutral 
absorption. However, we suggest that this is a fit artifact due to the presence
of an absorption feature at $\sim$0.5~keV, because of the flux increase 
observed by the LECS below $\sim$0.4~keV.

In previous ASCA observations of \src~
extra neutral absorption has been used as a description of possible 
SIS systematic uncertainties (see Cappi et al. 1998, Cappi \& Matsuoka 1997 
and references therein). 
Furthermore, because ASCA cannot observe the low-energy side of any
absorption feature at $\sim$0.5~keV, such  features can be mis-interpreted
as a systematic instrumental effect.
Simultaneous LECS/SIS power-law fits including excess absorption 
show that the 90\% confidence upper limits for ${\rm N_{sys}}$ relative to 
the LECS are $<$0.8$\times 10^{20}$ and $<$2.0$\times 10^{20}$ cm$^{-2}$ 
for SIS0 and SIS1, respectively in the range 0.6--4~keV. These upper
limits, together with the small {\it rms} spread in data-to-model ratio 
(see Sect.~\ref{systematic}) not only indicate a good match between LECS 
and SIS spectra, but exclude any other significant systematic trends with a 
spectral dependency. The variations in N$_{\rm H}$ amongst the 
different ASCA observations of \src~(Cappi \& Matsuoka 1997) are 
therefore likely to be
due to real changes in the intrinsic spectrum, rather than a measure of the
systematic error.  

The mean SIS/LECS normalization factor of $\sim$1.1
(see Table \ref{tabsystem} and Sect. \ref{absfits}) probably
reflects  inadequate modeling of the obscuration caused 
by the LECS strongback
during the off-axis exposure. The modeling is complicated by a slight
``wobble'' in pointing position during the BeppoSAX observation, 
due to a lack of suitable guide stars. 
%because the Z star-tracker of BeppoSAX cannot be used for \src. 
Future work is expected to improve the understanding of these effects.  

Strictly speaking the results of the present SIS/LECS comparison are 
only valid at the time of the observations and may not hold for other epochs.
Further simultaneous ASCA/BeppoSAX observations of \src~ are planned 
to monitor the ASCA and BeppoSAX intercalibration.
%performance of the SIS and the LECS.  

The SIS and the LECS have fundamentally different detectors
(CCDs versus GSPCs) and mirror designs (although the mirror 
materials are similar) and both instruments have been calibrated 
independently of one another. Therefore and because the relative agreement 
between the two instruments is good, one may be confident in the correctness
of their absolute calibration. 
For these reasons and because the common spectral range extends down to 
lower energies, the simultaneous fits to the SIS0/SIS1/LECS data bring  
the strongest constraints yet on the spectral shape of \src~ between
0.1--2~keV.

The photoionization models considered in Sect. \ref{photfits} require a 
large red-shift for the absorption component.
The relativistic in-flow velocity implied by such a red-shift 
(v$\sim$0.36~c) is higher than  the typical velocities found in the 
broad line regions of AGN
(1500--30000 km s$^{-1}$) or in broad absorption line systems, 
which don't exceed $\sim$0.1${\rm c}$ (Hamman 1997). There is some
evidence in other AGN for similar X-ray absorption features which may 
originate in relativistic winds (v$\sim$0.2--0.6 c, Leighly et al. 1997). 
However, it is not yet clear which mechanism accelerates these flows. 
Finally, the assumption of simple photoionization equilibrium may not 
apply and other physical interpretations of the absorption feature, such as
resonant absorption from O{\sc viii} Ly$\alpha$ at
E$_{rest} = 0.654$~keV, cannot be strictly excluded.

\begin{acknowledgements}
The authors wish to thank both the \asca\ and
 the \sax\ science team members for help with these
observations. We thank  O.R. Williams for
helpful comments. AO acknowledges an ESA Fellowship.
The \sax\ satellite is a joint Italian and Dutch programme.

\end{acknowledgements}


\begin{thebibliography}{}
\bibitem[1976]{}
Anders E., Grevesse N., 1989, Geochimica et Cosmochimica Acta 53, 197 
\bibitem[1976]{}
Boella G., Butler R.C., Perola G.C., et al., 1997a, A\&AS 122, 299
\bibitem[1976]{}
Boella G., Chiappetti L., Conti G., et al., 1997b, A\&AS 122, 327
\bibitem[1976]{}
Cappi M., Matsuoka M., 1997. In: Winkler C., Courvoisier T., Durouchoux P. 
(eds.) Proc. 2nd INTEGRAL Workshop, The Transparent Universe. ESA SP-382, 
p. 389 
\bibitem[1976]{}
Cappi M., Matsuoka M., Otani C., Leighly K., 1998, PASJ, in press
\bibitem[1976]{}
Courvoisier T.J-L, Turner M.J.L., Robson E.I., et al., 1987, A\&A 176, 197
\bibitem[1976]{}
Dotani T., Mitsuda K., Ezuka H., et al., 1996, ASCA News n.4 
\bibitem[1976]{}
Dotani T., Yamashita A., Ezuka H., et al., 1997, ASCA News n.5
\bibitem[1976]{}
Ferland G.J., 1996, University of Kentucky, Department of Physics and 
Astronomy, Internal Report
\bibitem[1976]{}
Fossati G., Haardt F., 1997, Internal report SISSA 146/97/A, astro-ph/9804282
\bibitem[1976]{}
Grandi P., Guainazzi M., Mineo T., et al., 1997, A\&A 325, L17
\bibitem[1976]{}
Grevesse N., Anders E., 1989, AIP Conference Proceedings 183, 
Ed. Waddington C., p. 1 
\bibitem[1976]{}
Hamman F., 1997, ApJS. 109, 279
%\bibitem[1976]{}
%Inoue H., Koyama K., Matsuoka M., Ohashi T., Tanaka Y., 1978, Nucl. 
%Instrum. \& Meth. A. 157, 295
\bibitem[1976]{}
Lammers U., 1997, The SAX LECS Data Analysis System, 1997, Internal Report
ESTEC SAX/LEDA/0010, ESA
\bibitem[1976]{}
Leach C., McHardy I., Papadakis I., 1995, MNRAS 272, 221
\bibitem[1976]{}
Leighly K., Mushotzky R., Nandra K., Forster K., 1997, ApJ 489, L25 
\bibitem[1976]{}
Morisson D., McCammon D., 1983, ApJ 270, 119
\bibitem[1976]{}
Orr A., Parmar A.N., Yaoqoob T., Guainazzi M., 1998, in Scarsi et al. (eds.)
Proc., The Active X-ray Sky: Results from BeppoSAX and Rossi-XTE, Rome,
October 1997. Nuclear Physics B Proceedings Supplements (in press),
astro-ph/9712274
\bibitem[1976]{}
Parmar A.N., Martin D.D.E., Bavdaz M., et al., 1997, A\&AS 122, 309
\bibitem[1976]{}
Savage B., Lu L., Bahcall J., et al., 1993, ApJ 413, 116
\bibitem[1976]{}
Serlemitsos P., Jalota L., Soong Y., et al., 1995, PASJ 47, 105
\bibitem[1976]{}
Stark A., Gamie C., Wilson R., et al., 1992, ApJS 79, 77 
\bibitem[1976]{}
Staubert R., 1992, In: W. Brinkmann and J. Tr\"umper (eds.)
X-ray Emission from Active Galactic Nuclei and the Cosmic X-ray Background,  
MPE report 235, p. 42 
\bibitem[1976]{}
Staubert R., Fink H., Courvoisier T.J-L. et al., 1992, In:
AIP Proc. 254, Testing the AGN Paradigm, Am. Inst. Phys., New York, p.366 
\bibitem[1976]{}
Tanaka Y., Inoue H., Holt S.S., 1994, PASJ 46, L37
\bibitem[1976]{}
Turner M.J.L., Williams O.R., Courvoisier T.J-L., et al., 1990, MNRAS 244, 310
\bibitem[1976]{}
Turner T.J., Weaver K.A., Mushotzky R.F., Holt S.S., Madejski G.M., 1991,
ApJ 381, 85
\bibitem[1976]{}
Walter R., Orr A., Courvoisier T.J.L., et al, 1994, A\&A 285, 119 
\bibitem[1976]{}
Williams O.R., Turner M.J.L., Stewart G.C., \etal, 1992, ApJ 389, 157
\bibitem[1976]{}
Worrall D., Mushotzky R., Boldt E., Holt S., Serlemitsos P., 
1979, ApJ 232, 683
\bibitem[1976]{}
Yaqoob T., Serlemitsos P., Mushotzky R., \etal, 1994, PASJ 46, L49
\bibitem[1976]{}
\end{thebibliography}
\end{document}